\documentclass[twocolumn,showpacs,aps,prl,superscriptaddress]{revtex4}

\usepackage{graphicx}
\usepackage{dcolumn}
\usepackage{epsfig}
\usepackage{amsmath}
\usepackage{units} 
\RequirePackage{xspace}
\usepackage{relsize} 

\newcommand{\etal}{{\it et al.}}

\newcommand{\btorhokz}{{B^+\to\rho^+K^0}}
\newcommand{\ach} {\ensuremath{ {\cal A}_{\rm ch} }\xspace}
\newcommand{\mrho} {\ensuremath{m_{\pip\piz} }\xspace}

\newcommand{\BABARPubYear}    {07}
\newcommand{\BABARPubNumber}  {005}

\newcommand{\SLACPubNumber} {12210}

\input pubboard/babarsym.tex

\begin{document}

\preprint{\babar-PUB-\BABARPubYear/\BABARPubNumber} 
\preprint{SLAC-PUB-\SLACPubNumber} 

\begin{flushleft}
\babar-PUB-\BABARPubYear/\BABARPubNumber \\
SLAC-PUB-\SLACPubNumber \\
\end{flushleft}

\title{ \large \bf\boldmath Observation of 
$B^+\to\rho^+\Kz$ and Measurement of its
Branching Fraction and Charge Asymmetry} 

%% author list as of 05-Jan-2007 (582 authors)
%
\author{B.~Aubert}
\author{M.~Bona}
\author{D.~Boutigny}
\author{Y.~Karyotakis}
\author{J.~P.~Lees}
\author{V.~Poireau}
\author{X.~Prudent}
\author{V.~Tisserand}
\author{A.~Zghiche}
\affiliation{Laboratoire de Physique des Particules, IN2P3/CNRS et Universit\'e de Savoie, F-74941 Annecy-Le-Vieux, France }
\author{J.~Garra~Tico}
\author{E.~Grauges}
\affiliation{Universitat de Barcelona, Facultat de Fisica, Departament ECM, E-08028 Barcelona, Spain }
% ...
\collaboration{The \babar\ Collaboration}
\noaffiliation

\begin{abstract}
We present the first observation of the decay $B^+\to\rho^+K^0$,
using a data sample of 348~fb$^{-1}$ collected at the $\Upsilon(4S)$
resonance with the \babar\ detector. 
The branching fraction and charge asymmetry are measured to be $(8.0^{+1.4}_{-1.3}\pm0.6)\times10^{-6}$ and ($-12.2\pm16.6\pm2.0)\%$, respectively,
 where the first uncertainty is statistical and the second is systematic. 
The charge asymmetry is defined by
$\ach=(\Gamma_{\Bm}-\Gamma_{\Bp})/(\Gamma_{\Bm}+\Gamma_{\Bp})$
with $\Gamma_{\Bpm}$ the \Bpm decay rate.
The significance of the observed branching fraction, including systematic
uncertainties, is 7.9~standard deviations.
\end{abstract}

\pacs{13.25.Hw, 12.15.Hh, 11.30.Er} 

\maketitle

In the Standard Model (SM) of particle physics,
the weak-current couplings of quarks are described by elements of the 
Cabibbo-Kobayashi-Maskawa (CKM) matrix~\cite{ckm}.
Charmless decays of \B mesons provide important information 
about these couplings.
These decays, which have branching fractions of the order of $10^{-6}$, 
are generally expected to occur via $b\to s$ or $b\to d$ virtual 
loop (``penguin")
amplitudes, tree-level $b\to u$ decays, or a combination of the two.
Phenomenological fits to the branching fractions and charge asymmetries
of charmless \B decays can be used to understand the relative importance
of tree and penguin amplitudes and to extract measurements
of the CKM phase angles.

We present  the first observation of 
the charmless \b\ra\s process $\Bp\to\rho^+\Kz$.  Throughout this paper,
the charge conjugate channel is implied unless otherwise stated.
We measure the branching fraction and charge asymmetry.
The latter is defined as 
$\ach=(\Gamma_{\Bm}-\Gamma_{\Bp})/(\Gamma_{\Bm}+\Gamma_{\Bp})$
with $\Gamma_{\Bpm}$ the \Bpm decay rate. A non-zero value of $\ach$ implies
violation of \CP symmetry.  Data were collected with the {\babar} detector
at the {\pep2} asymmetric \epem\ collider at the 
Stanford Linear Accelerator Center.
The data used in the analysis are based on a  sample with an 
integrated luminosity of 348~fb$^{-1}$, 
corresponding to $383\pm4$ million {\BB} pairs  
recorded at the $\Upsilon (4S)$ resonance 
[center-of-mass energy (CM) $\sqrt{s}=10.58\ \gev$]. 

The $\Bp\to\rho^+\Kz$ decay is expected to be a pure penguin decay~\cite{ChiangVP}, 
making it particularly helpful to separate the contributions
of tree and penguin amplitudes in other channels.
Phenomenological studies~\cite{bib-gronau-2000,ChiangVP,bib-chiang-2002}
 of charmless, strangeness changing ($|\Delta S|=1$) $\B\to VP$ decays, with $V$ a vector and
$P$ a pseudoscalar meson, assume that the penguin amplitudes $p'_V$ and
$p'_P$ are related by $p'_V=-p'_P$, where $p'_V$ ($p'_P$) is the amplitude
for the spectator quark to appear in the $V$ ($P$) meson. Measurement of
the $\Bp\to\rho^+\Kz$ branching fraction can provide a direct test of
this assumption~\cite{bib-lipkin}.  Exploiting $U$-spin symmetry,
Soni and Suprun~\cite{Uspin} recently introduced a technique 
to determine the CKM phase angle \g with precision
comparable to the best current measurements, 
using charmless $\Bpm\ra M^\pm M^0$ decays,
where $M^\pm$ and $M^0$ are charged and neutral mesons.
Of the eight $M^\pm M^0$ channels necessary to apply this technique
to $\Bpm\ra V^\pm P^0$ decays,
experimental results exist for all but two channels:
$\Bp\to\rho^+\Kz$, the topic of this study, and $K^{*+}\Kzb$.

Theoretical predictions of the branching fraction for $\Bp\to\rho^+\Kz$,
based on QCD factorization~\cite{QCDF,DU},
heavy quark effective theory~\cite{HQET},
and flavor SU(3) symmetry~\cite{ChiangVP,SU3P},  
vary from $10^{-5}$ to $10^{-6}$.  The only current experimental result is
${\cal B}(\Bp\to\rho^+\Kz)<4.8\times 10^{-5}$ at 90\% confidence level 
(CL)~\cite{bib-cleo-96}.
The charge asymmetry for this decay is expected to be zero. 
Any significant deviation from this expectation could 
provide evidence for the creation of non-SM particles
produced in the loops.

The {\babar} detector is described elsewhere~\cite{BABARNIM}.
In brief, charged particle tracks are detected and their momenta measured
by a combination of a five-layer double-sided
silicon microstrip detector (SVT) and a 40-layer drift chamber (DCH), 
both operating in the 1.5~T magnetic field of
a superconducting solenoid.
Tracks are identified as charged kaons or pions
using specific energy loss
measurements in the SVT and DCH as well as 
radiation angles measured in a ring imaging Cherenkov detector.
Photons are reconstructed from energy clusters deposited in
a CsI(Tl) electromagnetic calorimeter.

Monte Carlo (MC) events are used to determine
signal and background characteristics, 
optimize selection criteria, and evaluate efficiencies.
Samples of $\epem\ra\FourS\ra\BzBzb$ and \BpBm events,
generated by the \evtgen~\cite{bib-evtgen} event generator, 
 are passed through the \geant-based~\cite{GEANT} \babar\ detector 
simulation.  The number of MC events corresponds to about three times the
   integrated luminosity of the data.
We follow a blind procedure in which the optimization and
systematic study of selection criteria, and tests of the fitting 
procedure, described below, are completed before
the data are examined in the region where the signal is expected.

A $B$ meson candidate is kinematically characterized
by the beam-energy substituted mass 
$\mes\equiv\sqrt{s/4-(p_B^*)^2}$ and the energy difference
$\DeltaE\equiv E_B^*-\sqrt s/2$,
where $E_B^*$ and $p_B^*$ are the CM
energy and 3-momentum of the $B$ candidate, respectively.
Signal events peak at the nominal $B$ mass for \mes
and at zero for \DeltaE.

We reconstruct $\btorhokz$ candidates through the decays
$K^0\to\KS\to\pip\pim$ and $\rho^+\to\pip\piz$, 
with $\piz\to\g\g$.  The $\g$ energy in the laboratory frame is required
to exceed 30~MeV.  The \piz candidates are required to have a  mass in the
interval [0.115, 0.150]~\gevcc and a laboratory energy
larger than 0.2~GeV. The \piz mass resolution is about 6 MeV/$c^2$. 
To improve the resolution of $\mes$ and $\DeltaE$,
the \piz candidate's mass is constrained to its nominal value.
The \piz candidate is combined with an identified charged pion 
to form a $\rho^+$ candidate, which is required to  
have a mass $m_{\pi^+\pi^0}$  
in the interval [0.5, 1.0]~\gevcc.  The helicity angle $\theta_\rho$,
defined as the angle in the $\rho^+$ rest frame between the
direction of the boost from the \Bp rest frame and the 3-momentum of
the \pip from the $\rho^+$ decay,
is required to satisfy $|\cos\theta_\rho|<0.9$, since   
mis-reconstructed \piz mesons are concentrated
 near $|\cos(\theta_\rho)|\approx 1$.  We form \KS candidates 
by combining all oppositely charged pairs of tracks,
by fitting the two tracks to a common vertex,
and by requiring the mass to lie in
the interval [0.490, 0.506]~\gevcc assuming the two tracks to be pions.  
The \KS mass resolution is about 3 \mevcc. 
The angle $\alpha$ between the \KS flight direction and its momentum vector
is required to satisfy $\cos\alpha>0.995$, where the flight direction 
is the direction between the primary and secondary vertices.
The \KS candidate is combined with the $\rho^+$ candidate 
to form a \Bp candidate with a vertex constrained
 to the beam spot.  The \KS decay length significance,
defined as the ratio of the distance between the \KS and \Bp decay vertices
and the uncertainty on that quantity, is required to be larger than~5.
The $\chi^2$ probabilities of the fitted \KS and \Bp vertices
are each required to exceed~0.5\%.  $B^+$ candidates are required to satisfy 
$5.25<\mes<5.29$~\gevcc and $|\DeltaE|<0.20$~GeV.
The typical resolution for \mes (\DeltaE) is approximately 3.0~\mevcc
(30~MeV).
We find that 9.8\% of the events contain two or more $\Bp\ra\rho^+\Kz$ 
candidates.  These are mostly events with more than one reconstructed \piz.
 For these events, the candidate with the largest \B vertex 
fit probability is retained. 

Backgrounds arise primarily from random combinations 
of tracks and clusters in $\epem\to\qqbar$ ($q=u,d,s,c$) continuum events. 
To suppress these events, we follow a procedure similar to that
  described in Ref.~\cite{Fisher}.
We use the angle $\theta_T$ between the thrust axis of the $B$ candidate's 
decay products 
and the thrust axis determined using the remaining charged tracks and neutral
clusters in the event, evaluated in the CM frame.
 The distribution of $|\cos\theta_T|$ is nearly uniform for the 
almost-isotropic \BB events
and sharply peaked near 1 for the jetlike continuum events.  
We require $|\cos\theta_T|<0.9$. 
Additional use of the event topology is made by employing
a Fisher discriminant ${\cal F}$, constructed
from the angles with respect to the beam axis of the \B momentum
and the \B thrust axis, and the energy flow around the \B thrust 
axis~\cite{Fisher}.

Potential backgrounds from \BpBm and \BzBzb  events arise from 
$\Bb\to\Db\pi$, $K^{*}(892)\pi$ and $K_0^{*}(1430)\pi$ decays 
that have the same $\pi^+\pi^0 K_S^0$ final state 
and similar peaking structure in \mes and \DeltaE as the signal events. 
The selection requirement applied to \mrho, given above,
rejects most of these backgrounds.  To further reduce the \BB background,
we apply 
a \Dz veto ($1.78\le m_{\KS\piz}\le1.94$) \gevcc,
a \Dp veto ($1.83\le m_{\KS\pip}\le1.91$) \gevcc, 
a \Kstarz veto ($0.8\le m_{\KS\piz}\le1.0$) \gevcc, 
a \Kstarp veto  ($0.8\le m_{\KS\pip}\le1.0$) \gevcc, 
a $K^{*0}(1430)$ veto ($1.3\le m_{\KS\piz}\le1.6$) \gevcc, 
and a $K^{*+}(1430)$ veto ($1.3\le m_{\KS\pip}\le1.6$) \gevcc,
where a veto indicates that an event is rejected if  the two-particle 
invariant mass lies in the specified mass window. 
These veto criteria are determined as follows:
4 standard deviations of the experimental resolution around
the mass peaks for \Dp and \Dz, and 2 (1) resonance widths~\cite{PDG}
around the mass peak for \Kstar [$K^*(1430)$].
The decay $\Bp\to\KS\pip$ ($\Bz\to\KS\piz$) can contribute to
background when the decay products are combined with a low momentum 
$\piz$ ($\pip$).  We therefore require the
${\KS\piz}$ and ${\KS\pip}$ invariant masses
to be less than~5.2~\gevcc. 

 These criteria reject more than 99\% of the \BB background 
channels discussed above and about  30\% of the signal.
The remaining \BB background is 
combinatoric and does not peak in \DeltaE and \mes.

From MC simulation, the signal efficiency is determined to be 
($14.78\pm0.10$)\% where the uncertainty is statistical.
This efficiency has been corrected to account for small
differences in neutral particle reconstruction 
efficiencies between the data and MC, 
and for differences in the identification 
efficiency of the \pip used to reconstruct the~$\rho^+$. 
As an example, the latter correction is determined using a 
$\Dstarp\to\Dz\pip$ data control sample with $\Dz\to\Km\pip$.
The efficiency corrections are 97\% for the \piz reconstruction 
and greater than 99\% for the \KS reconstruction and \pip identification.

The number of signal events (the signal yield) and charge asymmetry
are determined from an extended unbinned maximum likelihood
(ML) fit with the following variables: 
\mes, \DeltaE, ${\cal F}$, $\mrho$, $\cos\theta_\rho$, 
and the \B flight time significance,
with the latter variable defined as the proper time difference $\Delta t$ 
between the produced \B and \Bb candidates divided by its 
uncertainty $\sigma_{\Delta t}$~\cite{JpsiKS}. 
The \Bb vertex is determined by fitting all tracks
except the daughters of the signal \B candidate to a common vertex, employing
constraints from the beam spot.  The likelihood function has the form 
\begin{equation}
  {\cal L}= {1\over N!}\exp{(-\sum_{j=1}^3 n_j)} \prod_{i=1}^N
  \left[\sum_{j=1}^3 n_j  {\cal P}_j ({\bf x}_i)\right],
\end{equation}
where $N$ is the total number of input events,
$n_j$ is the fitted yield of component $j$ 
(signal, continuum, and \BB background), 
and ${\cal P}_j ({\bf x}_i)$ is the corresponding  
overall probability density function (PDF), given by
\begin{eqnarray} 
{\cal P}_j&=&{\cal P}_j(\mes)\,{\cal P}_j(\DeltaE)\,
{\cal P}_j({\cal F})\,
{\cal P}_j(\mrho)\nonumber\\
&&\times\,
{\cal P}_j(\cos\theta_\rho)\,
{\cal P}_j(\Delta t/\sigma_{\Delta t}).
\label{PDFs6}
\end{eqnarray}
The signal and \BB background PDFs are determined from MC simulation. 
The continuum background PDF is obtained from sideband data 
($0.1<|\DeltaE|<0.2$ GeV for $\mes$ and $5.25<\mes<5.27$ GeV/c$^2$ for 
other variables).  For \mes, the PDFs of the signal and continuum are
parameterized by a Crystal Ball~\cite{CBALL}
and an ARGUS function~\cite{ARGUS}, respectively.
A relativistic Breit-Wigner function with a $p$-wave Blatt-Weisskopf
form factor~\cite{Blatt-W} is used to model the signal \mrho distribution.
For the background components, the \pip\piz mass is modeled by a combination of 
a polynomial and the signal function.  Slowly varying distributions 
($\DeltaE$ for the continuum background, 
and $\cos\theta_\rho$) are modeled by polynomials. 
The remaining variables are parameterized with either a Gaussian, 
the sum of two or three Gaussians, or an asymmetric Gaussian.
Dips occur near $|\cos\theta_\rho|=0.81$ because of the resonance vetoes.
We describe these dips by two Gaussian shapes.
We use a large data control sample of $\Bp\to \Dzb\pip$ 
($\Dzb\to K_S^0\pi^0$) events to verify the simulated 
resolutions and peak positions of the
$\mes$, $\DeltaE$, ${\cal F}$, and $\Delta t/\sigma_{\Delta t}$ 
signal PDFs.

Eq.~(\ref{PDFs6}) is based on the assumption that the variables
in the PDFs are uncorrelated. 
To evaluate possible bias in the signal yield that might arise 
from residual correlations, we construct an ensemble of 600
simulated experiments.  Each experiment contains the expected
number of signal, continuum, and \BB background events.  The
continuum events are randomly drawn from the PDFs while the signal and
\BB background events are randomly drawn from the MC samples.  The
bias is defined as the difference between the mean signal yield, determined
 from fits to the simulated experiments, and the number of signal
events included in the samples.
The bias in the signal yield is determined to be
$4.8\pm 1.2$ events, where the uncertainty is statistical.

\begin{table}[htbp]
\caption{Summary of results.
The uncertainties on the event yields, fit bias,
and efficiencies are statistical only.
}  
\begin{tabular}{lc}
\hline 
\hline 
Parameter & Value \\
\hline 
Events in fit         & 41150  \\ 
Signal yield (events) & $158^{+27}_{-26}$      \\ 
Continuum background yield (events) & 40,321$^{+210}_{-211}$ \\ 
\BB background yield (events)       & $673^{+71}_{-70}$ \\ 
Fit bias     (events) & 4.8$\pm$1.2            \\ 
Detection efficiency (\%)         & 14.78$\pm$0.10     \\  
Daughter branching fractions %& \\
$\prod {\cal B}_i$ (\%)  & 34.18$\pm$0.03     \\ 
Statistical significance ($\sigma$)        & 8.2 \\ 
Significance with systematics ($\sigma$)  & 7.9 \\ \hline  
Branching fraction ${\cal B}$ ($\times10^{-6}$)   & $8.0^{+1.4}_{-1.3}\pm0.6$\\ 
Charge asymmetry ${\cal A}_{\rm ch}$ (\%)    & $-12.2\pm16.6\pm2.0$ \\ 
\hline 
\hline 
\end{tabular} 
\label{tab:summary} 
\end{table}

\begin{figure}[thbp]
\center{\mbox{\includegraphics[width=0.50\textwidth]{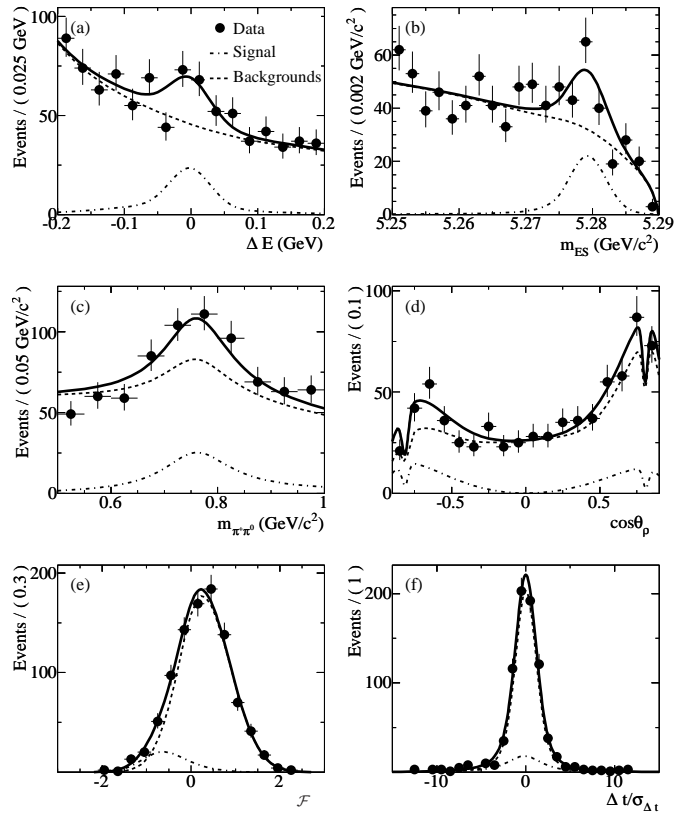}}}
\caption{
Distributions of (a)~\DeltaE, (b)~\mes,
(c)~\mrho, (d)~$\cos\theta_\rho$, (e)~${\cal F}$ , 
and (f)~$\Delta t/\sigma_{\Delta t}$. To improve visibility, 
a selection requirement on the likelihood ratio that 
retains 70.0\% of the signal events has been applied.
The points with uncertainties are the data.
The curves are projections of the ML fit.
The dashed curves show the sum of the continuum and
\BB background components.
The dot-dashed curves show the signal component.
The solid curves show the sum of the signal and background components. 
}
\label{projrhoks}
\end{figure}

Table~\ref{tab:summary} lists the results of the fit to the data. 
The fit yields a simultaneous determination of the fraction $f_{+}$ 
($f_{-}$) of \Bp (\Bm) events relative to the total number of 
signal events, 
with the constraint 
$f_{+}+f_{-}=1$.  
The charge asymmetry is determined from  
$\ach=2f_{-}-1$. 
The statistical uncertainty of the signal yield is given by 
the change in the central value when the quantity $-2\ln{\cal L}$ 
increases by one unit from its minimum value. 
The statistical significance is
given by the square root of the difference between the 
value of $-2\ln{\cal L}$ 
for zero signal events and the value at its minimum.
The corresponding significance including systematic uncertainties
(discussed below) is determined by convolution of the
likelihood function with a Gaussian distribution whose standard
deviation equals the total systematic uncertainty.
Figure~\ref{projrhoks} shows projections 
of the fitted variables.
To enhance the visibility of the signal,
events are required to satisfy 
${\cal L}_i(S)/[{\cal L}_i(S)+{\cal L}_i(B)]>0.9$  (this 
retains 70.0\%, 1.4\%, and 14.5\% of the signal, 
continuum, and \BB background events, respectively),
where ${\cal L}_i(S)$ is the likelihood function for signal events
excluding the PDF of the plotted variable~$i$ 
and ${\cal L}_i(B)$ is the corresponding sum for 
all background components. 

We calculate the branching fraction by subtracting
the fit bias from the measured signal yield
and  dividing the result by the overall efficiency and
the number of produced \BB pairs $N_{\BB}$.
The overall efficiency is the product of the detection efficiency 
and the daughter branching fractions~\cite{PDG} (see Table~\ref{tab:summary}).
We assume equal decay rates of the \FourS to $\Bp\Bm$ and~$\Bz\Bzb$. 
The branching fraction  and charge asymmetry 
 are determined to be $(8.0^{+1.4}_{-1.3}\pm0.6)\times10^{-6}$
and $(-12.2\pm16.6\pm2.0)\%$, respectively, where the first
    uncertainty is statistical and the second is systematic.  
We determine $-0.40<{\cal A}_{\rm ch}<0.15$ at 90\%~CL,
including systematic uncertainties.

The principal sources of systematic uncertainty are as follows.
The uncertainty in the \piz reconstruction efficiency is 3\%. 
The uncertainty related to the signal PDFs, 
assessed by varying the fitted PDF parameters within their 
uncertainties as determined from the $\Bp\to \Dzb\pip$ data 
control sample, is 3.2 events.  An uncertainty in the fit bias 
(2.7 events) is defined by the quadratic sum of half the bias 
itself and the statistical uncertainty of the bias. 
To evaluate the effect of a possible non-resonant component,
we generate a Monte Carlo sample using a 3-body Dalitz amplitude
event generator for $\Bp\to\pi^+\pi^0\KS$, including the $\Kstarb(892)$, 
$\overline {\kern -0.2em K}{}\xspace_{0}^{*}(1430)$, 
$\rho$ resonances and a non-resonant amplitude.
Re-performing the ML fit with signal PDFs determined from this
sample results in a 3.5\% increase in the signal yield,
which we take to be the systematic uncertainty.
Variations of all resonance vetoes yield an uncertainty of 3.1\%. 
When the requirement on $\cos\theta_\rho$ is varied,
the results change by 2.0\%.  
Other principal sources of uncertainty are those
from the track reconstruction efficiency (1.6\%), 
the \BB background PDFs (2.0 events),
$N_{\BB}$ (1.1\%), and variation of the selection
criteria on $|\cos\theta_T|$ (1.0\%).
We add all terms in quadrature
to obtain the total systematic uncertainty.

In summary, we present the first observation of the pure 
penguin $b\ra s$ decay process $B^+\to\rho^+K^0$.
The significance of the measured branching fraction is 7.9 standard deviations. 
Using the assumption $p_V'=-p_P'$~\cite{bib-lipkin}, the $\Bp\to\rho^+\Kz$
 branching fraction is predicted to lie between about 9 and 13$\times10^{-6}$,
    consistent with our measurement within the uncertainties. 
The measured charge asymmetry is consistent with the SM expectation of zero. 

We are grateful for the excellent luminosity and machine conditions
provided by our \pep2\ colleagues, 
and for the substantial dedicated effort from
the computing organizations that support \babar.
The collaborating institutions wish to thank 
SLAC for its support and kind hospitality. 
This work is supported by
DOE
and NSF (USA),
NSERC (Canada),
IHEP (China),
CEA and
CNRS-IN2P3
(France),
BMBF and DFG
(Germany),
INFN (Italy),
FOM (The Netherlands),
NFR (Norway),
MIST (Russia),
MEC (Spain), and
PPARC (United Kingdom). 
Individuals have received support from the
Marie Curie EIF (European Union) and
the A.~P.~Sloan Foundation.

\end{document}